\documentclass[prc,twocolumn,showpacs
]{revtex4-1}
\usepackage{amsmath,amssymb,epsfig,bm,mathrsfs,feynmp,color}
\usepackage{slashed}
\newcommand{\be}{\begin{equation}}
\newcommand{\ee}{\end{equation}}
\newcommand{\bea}{\begin{eqnarray}}
\newcommand{\eea}{\end{eqnarray}}
\newcommand{\half}{\frac1 2}

\newcommand{\aap}{Astron.\& Astrophys.}
\newcommand{\sun}{\odot}

\newcommand{\de}{\partial}

\newcommand{\vectau}{{\bm \tau}}
\newcommand{\vecrho}{{\bm \rho}}

\definecolor{red}{rgb}{0.8,0,0}
\definecolor{orange}{rgb}{0.8,0.2,0.0}
\definecolor{blue}{rgb}{0.3,0.0,0.8}

\begin{document}

\title{Equation of state of hypernuclear matter: impact of hyperon--scalar-meson couplings}

 \author{Giuseppe Colucci} 
\affiliation{Institute for Theoretical
   Physics, J.~W.~Goethe University, D-60438 Frankfurt-Main, Germany}

\author{Armen Sedrakian}
\affiliation{Institute for Theoretical
   Physics, J.~W.~Goethe University, D-60438 Frankfurt-Main, Germany}

\begin{abstract}
  We study the equation of state and composition of hypernuclear
  matter within a relativistic density functional theory with
  density-dependent couplings. The parameter space of
  hyperon--scalar-meson couplings is explored by allowing for mixing
  and breaking of SU(6) symmetry, while keeping the nucleonic couplings
  constant fixed.  The subset of equations of state, which corresponds
  to small values of hyperon--scalar-meson couplings allows for
  massive $M\le 2.25 M_{\odot}$ compact stars; the radii of
  hypernuclear stars are within the range 12--14 km. We also study the
  equation of state of hot neutrino-rich and neutrinoless hypernuclear
  matter and confirm that neutrinos stiffen the equation of state and
  dramatically change the composition of matter by keeping the
  fractions of charged leptons nearly independent of the density prior
  to the onset of neutrino transparency. We provide
  piecewise polytropic fits to six representative equations of state
  of hypernuclear matter, which are suitable for applications in
  numerical astrophysics.
 \end{abstract}

\maketitle
\section{Introduction}\label{sec:1}
The integral parameters of compact stars depend on their equation of
state (hereafter EoS) at high densities. Measurements of pulsar masses
in binaries provide the most valuable information on the underlying
EoS because these, being deduced from binary system parameters, are
model independent within a given theory of
gravity~\cite{2012arXiv1211.2457K}.  The recent discovery of the
pulsar J1614-2230 in a binary orbit with a white dwarf provided the
initial evidence for 2 $M_{\odot}$ compact stars. The mass of the
compact star, measured via the Shapiro delay, is 1.97$\pm$ 0.04 $M_{\odot}$
~\cite{2010Natur.467.1057M}.  The recent observation of a relativistic
binary consisting of a white dwarf and a pulsar (J0348$+$0432) in
optical and radio bands, respectively, provided another measurement, 
with similar accuracy and slightly larger mass
$2.01\pm0.04$ $M_{\odot}$~\cite{2013arXiv1304.6875A}. The fact that
the masses were measured by different methods strengthens the idea
that massive (2 $M_{\odot}$) compact stars exist in nature.  The
first of these observations spurred an intensive discussion of the
phase structure of dense matter, which is consistent with
the implied observational lower bound on the maximum mass of any
sequence of compact stars~\cite{2010Natur.467.1057Mi,2010arXiv1012.2919S,2011EL.....9411002V,2011PhRvC..84c5809R,Bonanno:2011ch,2011A&A...536A..92B,2012NuPhA.881...62W,2011arXiv1112.6430L,2012PhRvC..85f5802W,2012PhRvC..85e5806O,2012arXiv1207.4872M,2013ApJ...764...12M,
2012ApJ...759...57L,2013PhRvC..87a5804D,2012PhRvC..86b5805G,
2012ApJ...756...56J,2012arXiv1211.1231Z,2012arXiv1212.1388C,2012arXiv1212.3686J,2012arXiv1212.5911P,2012arXiv1212.6803M,2013arXiv1301.0390G,2013arXiv1301.2675S,2013arXiv1302.3776G}.

Pulsar radii have been extracted, e.g., from modeling the x-ray
binaries under certain reasonable model assumptions, but the
uncertainties are
large~\cite{2011ApJ...732...88G,2010ApJ...719.1807G,2012ApJ...747...75G,2013ApJ...762...96B}.
An example that we use in our analysis, is the pulse
phase-resolved x-ray spectroscopy of PSR J0437-4715, which sets a
lower limit on the radius of a 1.76\, $M_{\odot}$ solar mass compact
star $R > 11$ km within 3$\sigma$ error~\cite{2013ApJ...762...96B}.

Large masses and radii are  evidence for the relative stiffness of
the EoS of dense matter at high densities. Large densities may require
substantial population of heavy baryons (hyperons), because these
become energetically favorable once the Fermi energy of neutrons
becomes of the order of their rest mass. Their onset then reduces
the degeneracy pressure of a cold thermodynamic ensemble and softens
effectively the EoS of dense matter. This decreases the maximum mass of a
compact stars to values  which contradict the observation of massive
compact stars in nature. This controversy between the theory and
observations is the essence of the ``hyperonization puzzle'' in
compact stars.

Although the emergence of heavy baryons (mainly $\Sigma^{\pm}$ and
$\Lambda$ hyperons) were considered even before the discovery of
pulsars and their identification with the neutron
stars~\cite{1960AZh....37..193A}, their existence in the cores of
neutron stars is still elusive. Seminal work on interacting
hyperonic matter was carried early after the discovery of
pulsars~\cite{1974NuPhA.230....1B,1974PhRvD...9.1613M}.  Systematic
studies of interacting dense hypernuclear matter were carried with the
advent of relativistic density functional
methods~\cite{1982PhLB..114..392G,1985ApJ...293..470G,1987ZPhyA.326...57G,1987ZPhyA.327..295G,1989NuPhA.493..549W,1990PhRvL..64...13K,1991NuPhB.348..345E,1991PhRvL..67.2414G,1995PhLB..349...11E,1996PhRvC..53.1416S};
for reviews see Refs.~\cite{weber_book,2007PrPNP..58..168S}. Most of 
these works predicted masses that are not much larger than the
canonical mass of a neutron star (in contradiction with modern
observations).  Masses on the order of $\le 1.8 M_{\sun}$ were
obtained later in non-relativistic phenomenological
models~\cite{1999ApJS..121..515B}. Fully
microscopic models based on hyperon-nucleon potentials, which include
the repulsive three-body forces, predict low maximal masses for
hypernuclear stars~\cite{2011PhRvC..84c5801S,2012JPhCS.342a2006L}.

Finite-temperature hyperonic matter in the presence of  a thermal bath of
neutrinos may differ significantly from the cold hypernuclear matter
found in evolved neutron stars. Previous studies
~\cite{1977SvA....21..727V,1999ApJ...513..780P,2004IJMPD..13..215B,2011PhRvC..83b5804B} 
showed that as the EoS becomes stiffer, the deleptonizing effect of charged
hyperons at zero temperature is removed in the presence of neutrinos
and the sequence in which hyperons appear changes. 

In this work we contribute to the investigation of the ``hyperonization puzzle''
by studying the parameter space of the coupling of hyperons to the
light scalar octet of mesons. The parameter space is partially
motivated by the hypernuclear potentials as well as investigations of
the  external field QCD sum rules that involve SU(6) symmetry
breaking and mixing effects~\cite{timmermans_scalar,timmermans_vector}.
 Because the structure and even content of
the scalar mesons is still a subject of discussion it appears
necessary to investigate this particular domain of hypernuclear
physics with respect to the consistency with the compact star
observations. Conversely, as we show below, the considered parameter
space can in fact account for massive hyperonic compact stars, thus
providing a possible solution to the ``hyperonization puzzle''.

To study the EoS of hypernuclear matter we use a model based on
relativistic density functional theory of Walecka and a 
density-dependent parametrization of the nucleon--meson couplings of
Ref.~\cite{ring2005}.  The EoS and composition of matter are
studied both in the zero-temperature limit relevant to mature compact
stars as well as at finite temperatures relevant for the hot
proto-neutron star stage of evolution. In the latter case we assume
neutrino-rich matter in $\beta$ equilibrium and explore the effects of
finite temperature and neutrino content on the stiffness and
composition of matter.  We confirm the general features found in the
previous analysis~\cite{1977SvA....21..727V,1999ApJ...513..780P,2004IJMPD..13..215B,2011PhRvC..83b5804B}
 and provide a detailed account of these effects within
our model.

This paper is structured as follows. In Sec.~\ref{sec:2} we discuss
the theoretical set-up of the relativistic density functional
theory. The parametrization of the coupling constant of the theory is
discussed in Sec.~\ref{sec:3}. Section~\ref{sec:4} is devoted to the
discussion of the choice of the hyperon-meson coupling constants.  We
present our results in Sec.~\ref{sec:5}. Finally, our conclusions are
collected in Sec.~\ref{sec:6}.

\section{Theoretical model}\label{sec:2}

The relevant degrees of freedom in nuclear matter at high density are
nucleons and hyperons.  In particular, we choose as degrees of freedom
of our relativistic mean field model the baryon octet ($J^P =
\half^+$), and we study the interaction of these fields with the
isoscalar-scalar, $\sigma$, isoscalar-vector, $\omega_\mu$, and
isovector-vector, $\vecrho_\mu$, mesons. Besides these fields, we also
consider the presence of leptons, $e^-,\mu^-$. In the case of 
finite-temperature matter, which describes proto-neutron star matter, the
model has the neutrino degrees of freedom, because they are expected
to be trapped in a proto-neutron star after the first minute of its
birth. The introduction of additional variables, the neutrino chemical
potentials, requires additional constraints, which we provide by fixing
the lepton fractions, $Y_{Ll}$, appropriate for conditions prevailing
in the evolution of a proto-neutron star \cite{1999ApJ...513..780P}.

The relativistic Lagrangian reads 
\bea\label{eq:lagrangian} \nonumber
{\cal L} & = & \sum_B\bar\psi_B\bigg[\gamma^\mu
\left(i\de_\mu-g_{\omega B}\omega_\mu 
- \half g_{\rho B}\vectau\cdot\vecrho_\mu\right)\\
\nonumber
& - & (m_B - g_{\sigma B}\sigma)\bigg]\psi_B 
+ \half \de^\mu\sigma\de_\mu\sigma-\half m_\sigma^2\sigma^2\\
\nonumber & - & \frac{1}{4}\omega^{\mu\nu}\omega_{\mu\nu} + \half
m_\omega^2\omega^\mu\omega_\mu
- \frac{1}{4}\vecrho^{\mu\nu}\vecrho_{\mu\nu} 
+ \half m_\rho^2\vecrho^\mu\cdot\vecrho_\mu\\
& + & \sum_{\lambda}\bar\psi_\lambda(i\gamma^\mu\de_\mu -
m_\lambda)\psi_\lambda - \frac{1}{4}F^{\mu\nu}F_{\mu\nu}, 
\eea 
where the $B$ sum is over the $J^P = \half^+$ baryon octet and  $\psi_B$ are the
baryonic Dirac fields with masses $m_B$. The interaction is mediated
by the $\sigma,\omega_\mu$ and $\vecrho_\mu$ meson
fields with the $\omega_{\mu\nu}$ and $\vecrho_{\mu\nu}$ field
strength tensors and masses $m_{\sigma}$, $m_{\omega}$, and
$m_{\rho}$.  The baryon-meson coupling constants are denoted by
$g_{mB}$ (their numerical value is discussed in Sec. \ref{sec:3}
below.)  The $\lambda$-sum in Eq. (\ref{eq:lagrangian}) runs over the
leptons $e^-,\mu^-,\nu_e$ and $\nu_\mu$ with masses $m_\lambda$ and
the last term is the electromagnetic energy density. The contribution
of neutrinos to the sum above is included only at non-zero
temperature, when they are trapped in stellar matter and form a
statistical ensemble in equilibrium. In mature compact stars at
essentially zero temperature the neutrinos do not contribute to the
thermodynamical quantities of matter because they are out of equilibrium
and their chemical potential vanishes.

The Lagrangian density (\ref{eq:lagrangian}) yields the pressure and
the energy density, which at finite temperature read respectively
\bea\label{eq:thermodynamic} 
P & = & - \frac{m_\sigma^2}{2} \sigma^2 +
\frac{m_\omega^2}{2} \omega_0^2 + \frac{m_\rho^2 }{2} \rho_{03}^2
+ \frac{1}{3}\sum_B \frac{2J_B+1}{2\pi^2} \nonumber\\
&\times &\int_0^{\infty}\!\!\! \frac{dk\ k^4 }{E_k^B}
\left[f(E_k^B-\mu_B^*)+f(E_k^B+\mu_B^*)\right]\nonumber\\
& +&\frac{1}{3\pi^2} \sum_{\lambda} \int_0^{\infty}\!\!\!  \frac{dk\ k^4 }{E_k^\lambda}
\left[f(E_k^{\lambda}-\mu_\lambda)+f(E_k^{\lambda}+\mu_\lambda)\right]\nonumber\\
\eea 
and
\bea {\epsilon} &=& \frac{m_\sigma^2}{2} \sigma^2 +
\frac{m_\omega^2 }{2} \omega_0^2 + \frac{m_\rho^2}{2} \rho_{03}^2
+ \sum_B \frac{2J_B+1}{2\pi^2}  \nonumber\\
&\times&\int_0^{\infty}dk \ k^2 E_k^B
\left[f(E_k^B-\mu_B^*)
+f(E_k^B+\mu_B^*)\right] \nonumber\\
&+& \sum_{\lambda} \int_0^{\infty} \!\!\!\frac{dk}{\pi^2}
k^2E_k^\lambda\left[f(E_k^\lambda-\mu_\lambda)
+f(E_k^\lambda+\mu_\lambda)\right],
\nonumber\\
\eea 
where $\sigma$, $\omega_0$ and $\rho_{03}$ are the nonvanishing mesonic mean
fields, $J_B$ is the baryon degeneracy factor, $m_B^* = m_B -
g_{\sigma B}$ is the effective baryon mass, $\mu^*_B =
\mu_B-g_{\omega B}\omega_0 - g_{\rho B}I_3\rho^0_3$ is the
baryon chemical potential including the time component of the fermion
self-energy, $I_3$ is the third component of baryon isospin,
$E_k^B = \sqrt{k^2+m^{*2}_B}$ and $E_k^{\lambda}=\sqrt{k^2+m_\lambda^2}$  are
the single particle energies of baryons and leptons respectively, and 
$f(y) = [1+\exp(y/T)]^{-1}$ is the Fermi distribution
function with $T$ being the temperature. We take each 
lepton mass  $m_\lambda$ equal to its free-space value.

At nonzero temperature the net entropy of the matter is the sum of the
baryon, $S_B$, and  lepton, $S_L$, contributions
\bea 
S_B &=& - \sum_B\frac{2J_B+1}{2\pi^2}\int_0^{\infty} dk k^2 
\Bigg\{\Bigl[f(E_k^B-\mu_B^*)\ln f(E_k^B-\mu_B^*) \nonumber\\
&+& \bar f(E_k^B-\mu_B^*)\ln \bar f(E_k^B-\mu_B^*) \Bigr] + (\mu_B^* \to -\mu_B^*)
\Biggr\}
\eea
and 
\bea
S_L &=&- \sum_{\lambda} \int_0^{\infty} \!\!\!\frac{dk}{\pi^2}
\Bigl[f(E_k^\lambda-\mu_{\lambda})\ln f(E_k^\lambda-\mu_{\lambda}) \nonumber\\
&+& \bar f(E_k^\lambda-\mu_{\lambda})\ln \bar f(E_k^\lambda-\mu_{\lambda})\Bigr],
\eea
where $\bar f(y) = 1-f(y)$. The free-energy density is then given by 
\be 
F  = \epsilon - T (S_B+S_L).
\ee

\section{Density-dependent parametrization DD-ME2}\label{sec:3}

Below, we work with a density-dependent parametrization, which
is designed to account for in an economical manner the many-body
correlations that arise beyond the mean-field approximation. The
density dependence of the coupling constant thus accounts for the
influence of the medium on the scattering of baryons. We choose to
work with the Density Dependent Meson Exchange 
(DD-ME2) parametrization of Ref.~\cite{ring2005}, which
introduces an explicit density dependence in the nucleon-meson
couplings $g_{N\sigma}$, $g_{N\omega}$ and $g_{N\rho}$.  The
phenomenological ansatz for the density dependence for the $\sigma$-
and $\omega_\mu$-meson coupling constants
is~\cite{typel:1999,ring2005} \be g_{iN}(\rho_B) =
g_{iN}(\rho_{0})h_i(x), \qquad i=\sigma,\omega, \ee where $\rho_B$ is
the baryon density, \be\label{ansatz} h_i(x) =
a_i\frac{1+b_i(x+d_i)^2}{1+c_i(x+d_i)^2} \ee is a function of
$x=\rho_B/\rho_{0}$ and $\rho_{0}$ is the nuclear saturation density. The
parameters in Eq.~(\ref{ansatz}) are not independent. Indeed, the five
constraints
$h_i(1)=1,h^{\prime\prime}_\sigma(1)=h^{\prime\prime}_\omega(1),h^{\prime\prime}_i(0)=0$
reduce the number of independent parameters to three. Three additional
parameters in the isoscalar-scalar channel are
$g_{N\sigma}(\rho_{0})$, $g_{N\omega}(\rho_{0})$ and $m_\sigma$, the
mass of the phenomenological $\sigma$ meson. Microscopic calculations
show that the $\vecrho_\mu$-meson coupling decreases at high
densities~\cite{typel:1999}. Therefore, instead of the
ansatz~(\ref{ansatz}) one can use a minimalistic ansatz, with
exponential form of density dependence~\cite{typel:1999,ring2005} \be
g_{\rho N}(\rho_B) = g_{\rho N}(\rho_{0})\exp[-a_\rho(x-1)].  \ee
Thus, the parametrization we use has in total eight
parameters, which are adjusted to reproduce the properties of
symmetric and asymmetric nuclear matter, binding energies, charge
radii, and neutron radii of spherical nuclei. The parameters of the
DD-ME2 effective interaction are shown in Table \ref{table:DD-ME2}.

\begin{table}[ht]
  \centering
\caption{Meson masses and couplings to the baryons in DD-ME2 effective interaction.}
\label{table:DD-ME2}
  \begin{tabular}{cccc}
\hline
           &     $ \sigma$    & $\omega$ & $\rho$  \\
\hline
    $m_i$ [MeV] &  550.1238  &  783.0000    &  763.0000  \\
    $g_{Ni}(\rho_{0})$ & 10.5396    & 13.0189     &  3.6836\\
    $a_i$  & 1.3881    & 1.3892 &   0.5647\\
    $b_i$  & 1.0943   & 0.9240 &   ---\\
    $c_i$  & 1.7057   &  1.4620 &   ---\\
    $d_i$  & 0.4421  & 0.4775 &    ---\\
\hline
  \end{tabular}
\end{table}
When the coupling constants are density dependent, the thermodynamical
consistency (specifically the energy conservation and fulfillment of
the Hugenholtz--van Hove theorem) require the inclusion of the so-called
rearrangement self-energy~\cite{1995PhRvC..52.3043F}. This contributes to the
pressure, but not to the energy of the system. The pressure becomes
\bea\label{eq:thermodynamic_p} P_r& = & P + \rho_B\Sigma_{r}, \eea
where the rearrangement self-energy, $\Sigma_{r}$, is given by
\be\label{eq:rearrangement} \Sigma_{r} = \frac{\partial
  g_{N\omega}}{\partial \rho_B}\omega_0\rho_B - \frac{\partial
  g_{N\sigma}}{\partial \rho_B}\sigma\rho_S, \ee where $\rho_S$ is
the scalar density. It can be verified that the contribution
from the rearrangement self-energy restores the thermodynamical
relation 
\be 
P_r = \rho_B^2 \frac{\partial}{\partial\rho_B}\left(\frac{\epsilon}{\rho_B}\right), 
\ee 
which is otherwise violated. Below we always consider the
pressure including the rearrangement term and drop the subscript
$r$.

\section{Hyperon-meson coupling constants}\label{sec:4}

To extend the description of matter to full baryon octet we need the
hyperon-meson coupling constants.  Because the information on the
properties of hypernuclear matter is far less extensive than for
nucleons their values can not be fixed with certainty and several
approaches have been used in the literature.

One way to fix these couplings is to consider the SU(3)-flavor
symmetric model, namely the octet model.  In this model the degrees of
freedom are represented by the lowest non-trivial irreducible
representation (IR) of the symmetry group which is physically
possible, $\{8\}$, that is the baryon octet and the mesonic octet.

Let us consider first the interaction between the baryon octet,
$J^P=\half^+$, and the vector meson octet, $J^P=1^-$.  Within the
assumption of SU(3)-flavor symmetry, one can express the
meson-baryon coupling constants in terms of only two parameters
\cite{de_swart1963}, the nucleon--$\vecrho_\mu$ meson coupling constant $g_{N\rho}$ and
the $F/(F+D)$ ratio of the vector octet $\alpha_V$:
\be\label{vector_relations}
\begin{split}
  &g_{N\rho} = g_8, \qquad\qquad  g_{\Xi\rho} = -g_8(1-2\alpha_V),\\
  &g_{\Sigma\rho} ~= 2 g_8 \alpha_V, \qquad g_{\Lambda\rho} = 0,\\
  &g_{N\omega_8} = {\frac{1}{\sqrt{3}}}g_8(4\alpha_V-1), 
\quad g_{\Xi\omega_8} = -{\frac{1}{\sqrt{3}}}g_8(1+2\alpha_V),\\
  &g_{\Sigma\omega_8} ~= {\frac{2}{\sqrt{3}}} g_8 (1-\alpha_V),
  \quad~ g_{\Lambda\omega_8} = -{\frac{2}{\sqrt{3}}} g_8,
  (1-\alpha_V),
\end{split}
\ee
where we have shown only the coupling constants relevant for our model. 
The vector-meson dominance model \cite{VMD_model} predicts
$\alpha_V=1$, which is a result of the universal coupling of the
$\vecrho_\mu$ meson to the isospin current. This result is also in
agreement with recent QCD sum rules calculations
\cite{timmermans_vector}.  Therefore, we set $\alpha_V = 1$ in
hyperon--$\vecrho_\mu$-meson coupling constants in
Eq.~(\ref{vector_relations}), i.e., 
\be \label{rho_couplings}
 g_{\Xi\rho} = g_8, \qquad g_{\Sigma\rho}= 2g_8, 
\qquad g_{\Lambda\rho} = 0.
\ee
To describe the mixing between the singlet and the octet members of
the vector nonet, one has to introduce two additional parameters, the
flavor singlet coupling constant, $g_1$, and the vector mixing angle,
$\theta_V$.  Therefore, the coupling of the baryon to the physical
$\omega_\mu$-meson state now reads 
\be\label{vector_mixing}
g_{B\omega} = \cos\theta_V g_1 + \sin\theta_Vg_{B\omega_8}.\ee
 It
is widely accepted that the mixing between nonstrange and strange
quark wave functions in the $\omega$ and $\phi$ mesons is ideal
\cite{2009JHEP...07..105K,chpt2006}. In this limit $\tan\theta_V = 1/\sqrt{2}$ and
assuming again $\alpha_V = 1$ we obtain the baryon--$\omega_\mu$ meson
couplings as
\be
\begin{split}
\label{eq:couplings81}
&\frac{g_{N\omega}}{\sin\theta_V}= \sqrt{2}g_1 + \sqrt{3}g_8,\\
&\frac{g_{\Xi\omega}}{\sin\theta_V} =  \sqrt{2}g_1 - \sqrt{3}g_8,\\
&\frac{g_{\Sigma\omega}}{\sin\theta_V}= \frac{g_{\Lambda\omega}}{\sin\theta_V} = \sqrt{2} g_1 .
\end{split}
\ee 
Next we assume that the nucleon does not couple to pure strange
mesons ($\phi_\mu$). One then obtains the relation between the singlet and
the octet coupling constants \cite{dover1984}: 
\be
g_1 = \sqrt{6}\,g_8,
\ee 
which allows one to eliminate $g_1$ from Eqs.~(\ref{eq:couplings81}).
Thus,  the hyperon--$\omega_\mu$-meson
coupling constants can be related to the nucleon one
\be\label{omega_couplings}
g_{\Xi\omega} = \frac{1}{3}g_{N\omega},\quad
g_{\Sigma\omega} = g_{\Lambda\omega} = \frac{2}{3}g_{N\omega}.
\ee
We are now left with the hyperon--scalar-meson ($\sigma$) coupling
constants.  In order to fix these parameters, we consider the
relations among the coupling
constants for the scalar octet~\cite{timmermans_scalar},
which are given  [in analogy with the relations for the vector meson
octet in Eq.~(\ref{vector_relations})] by  
\be\label{scalar_relations}
\begin{split}
&g_{Na_0} = g_S, \qquad\qquad g_{\Xi a_0} = -g_S(1-2\alpha_S),\\
&g_{\Sigma a_0} ~= 2 g_S \alpha_S, \qquad g_{\Lambda a_0} = 0,\\
&g_{N\sigma_8} = {\frac{1}{\sqrt{3}}}g_S(4\alpha_S-1), \quad
g_{\Xi\sigma_8} = -{\frac{1}{\sqrt{3}}}g_S(1+2\alpha_S),\\
&g_{\Sigma\sigma_8} ~= {\frac{2}{\sqrt{3}}} g_S (1-\alpha_S), \quad~ g_{\Lambda\sigma_8} = -{\frac{2}{\sqrt{3}}} g_S (1-\alpha_S).
\end{split}
\ee The mixing between the singlet and the octet states of the scalar
nonet is given by \be\label{scalar_mixing} g_{B\sigma} = \cos\theta_S
g_1 + \sin\theta_Sg_{B\sigma_8}.  \ee Then, the $\sigma$-meson--baryon
coupling constants are
\bea
g_{N\sigma}             & = & \cos\theta_S~g_1 +  \sin\theta_S~(4\alpha_S-1)g_S/\sqrt{3},\\
g_{\Lambda\sigma} & = & \cos\theta_S~g_1 - 2\sin\theta_S~(1-\alpha_S)g_S/\sqrt{3},\\
g_{\Sigma\sigma}   & = & \cos\theta_S~g_1 + 2\sin\theta_S~(1-\alpha_S)g_S/\sqrt{3},\\
g_{\Xi\sigma} & = & \cos\theta_S~g_1 -
\sin\theta_S~(1+2\alpha_S)g_S/\sqrt{3}.  \eea 
Combining the couplings, we obtain 
\bea 
g_{\Lambda\sigma} + g_{\Sigma\sigma}
& = & 2\cos\theta_S~g_1,\\
g_{N\sigma} + g_{\Xi\sigma}
& = & 2\cos\theta_S~g_1 - 2\sin\theta_S~(1-\alpha_S)g_S/\sqrt{3}\nonumber\\
& = & \cos\theta_S~g_1 + g_{\Lambda\sigma} .  
\eea 
By eliminating $g_1$ from these expressions we obtain a relation
\be\label{coupling_variation}
2(g_{N\sigma} + g_{\Xi\sigma}) = 3g_{\Lambda\sigma} +
g_{\Sigma\sigma}, 
\ee 
which is valid for arbitrary values of the four parameters $\alpha_S$,
$g_1$, $g_S$ and $\theta_S$. In particular, it is satisfied for the
values of the coupling constants in the SU(6) symmetric quark
model. Indeed, the latter model assumes that the $\sigma$ meson is a
pure up and down state, therefore for strangeness $-1$ $\Lambda$ and
$\Sigma$ baryons 
$g_{\Lambda\sigma}=g_{\Sigma\sigma}=\frac 2 3
g_{N\sigma}$, whereas for strangeness $-2$ $\Xi$ baryons
 $g_{\Xi\sigma}=\frac 1 3 g_{N\sigma}$. 
 Due to the constraint (\ref{coupling_variation}) the parameter space
 of our model is spanned by three out of the four parameters
 $\alpha_S$, $g_1$, $g_S$ and $\theta_S$. Further constraints on the
 parameter space can be placed because the hyperon coupling
 constants must be positive and less than the nucleon coupling
 constant. This implies two additional constraints, which can be
 translated into constraints on the range of variability of the
 hyperon--scalar-meson coupling constants. Indeed, by expressing one
 of the hyperon coupling constants, say $g_{\Xi\sigma}$, in terms of
 the others with the help of Eq.~(\ref{coupling_variation}) one finds
 \be \label{coupling_variation2} 
g_{\Xi\sigma} = \half
 (3g_{\Lambda\sigma} + g_{\Sigma\sigma}) - g_{N\sigma},
 \ee 
and therefore, by requiring that 
\be
\label{additional_constraints} 
0\le g_{\Xi\sigma}\le g_{N\sigma} , \ee one is left with the following
two simultaneous inequalities \be \label{additional_constraints2}
g_{N\sigma}\le \half (3g_{\Lambda\sigma} + g_{\Sigma\sigma}) \le
2g_{N\sigma}.  \ee To explore the parameter space spanned by the
hyperon-$\sigma$ meson couplings we use our nuclear density functional
and fix the remaining couplings. We fix the value of
$g_{\Lambda\sigma}$ at the value provided by the Nijmegen soft core
(NSC) hypernuclear potential model of Ref.~\cite{1989PhRvC..40.2226M}
and vary the range of couplings $g_{\Sigma\sigma}$ within the limits
provided by Eq. (\ref{additional_constraints2}). The corresponding
$g_{\Xi\sigma}$ couplings are found from
Eq.~(\ref{coupling_variation2}). Analogously, to explore the range of
admissible $g_{\Lambda\sigma}$ couplings, we next fix the value of
$g_{\Sigma\sigma}$ to the value provided by the NSC model and change
the $g_{\Lambda\sigma}$ coupling within a range which keeps the values
of $g_{\Xi\sigma}$ consistent with
Eq.~(\ref{additional_constraints}). The parameter space is further
limited by the requirement that the maximal values of
$g_{\Lambda\sigma}$ and $g_{\Sigma\sigma}$ should be below their
values in the SU(6) symmetric model. This latter constraint is
motivated physically by our search for a stiff EoS of hypernuclear
matter.

\section{Results}\label{sec:5}

The matter in evolved compact stars is charge neutral and in
$\beta$-equilibrium. These two conditions are imposed when computing
the EoS at zero temperature. For young neutron stars at
non-zero temperature a thermal ensemble of neutrinos should be
included for temperatures $T\ge 5$ MeV. In this case a common practice
is to vary the lepton fraction of matter in a certain range compatible
with astrophysical simulations. We treat these distinct
cases in turn below.

\subsection{Zero temperature}
\begin{figure}[t]
 \centering
\vspace{0.5cm}
 \includegraphics[width=7.5cm]{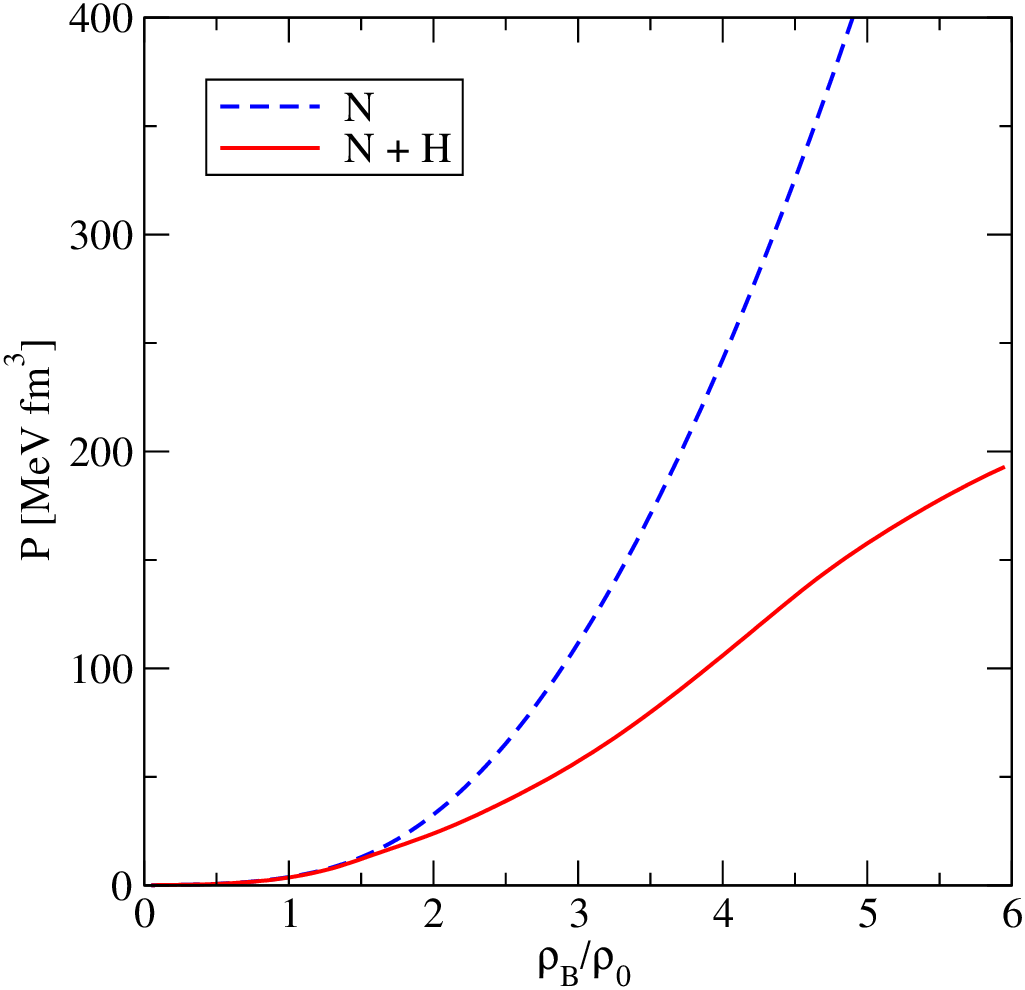}
 \caption{(Color online)
Equations of state of nuclear (dashed, blue online) and hypernuclear
(solid, red online) matter. The coupling constants for nuclear matter
correspond to density-dependent  (DD) ME2 parametrization of
Ref.~\cite{ring2005}.  The coupling constants of hyperons are related
to the nucleonic ones according to the SU(6) symmetric quark model.
The density is normalized to the nuclear saturation density $\rho_0 =
0.152$ fm$^{-3}$ in the DD ME2 parametrization.
}
 \label{fig:nucleonic}
\end{figure}

We start the discussion of our results by comparing the EoS of purely
nucleonic matter, computed using DD
parametrization of nuclear matter, to the EoS where interacting
hyperons are added with couplings taken according to the SU(6) quark
model; see Fig.~\ref{fig:nucleonic}. The appearance of hyperons, which
is triggered by the fact that the cost of having a hyperon heavier than
a nucleon is energetically more favorable than a neutron at the
top of the Fermi sea, always softens the EoS. Because in
our search the coupling constants are bound from above by their values
in the SU(6) model, these two EoS correspond to the stiffest and
softest EoS in our collection.

Our next step is to explore the impact of the variation of the 
hyperon--scalar-meson coupling on the EoS. As explained previously in
Sec.~\ref{sec:4}, we keep the hyperon--vector-meson couplings
fixed at the values given by Eqs.~(\ref{rho_couplings}) and
(\ref{omega_couplings}) and vary the hyperon--scalar-meson coupling
constants in the range and manner defined above.

\begin{figure}[t]
 \centering
\vspace{0.5cm}
 \includegraphics[width=8.5cm]{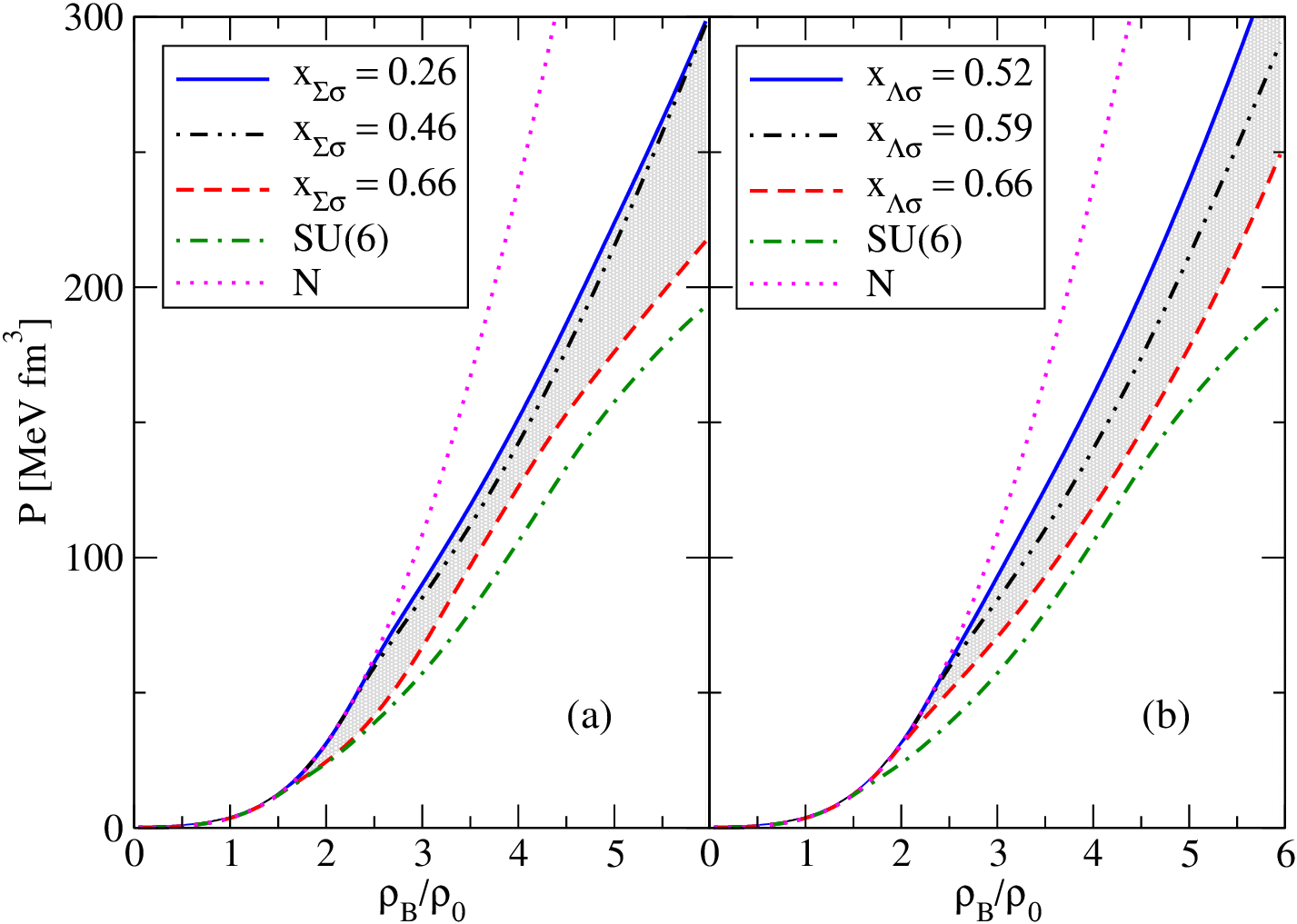}
 \caption{(Color online) Equations of state of hypernuclear matter for
   a range of values of hyperon--$\sigma$-meson couplings defined in
   terms of $x_{H\sigma} = g_{H\sigma}/g_{N\sigma} $, $H\in \Lambda,
   \Sigma$.  The nucleonic EoS (dotted line, magenta online) and
   hyperonic EoS with SU(6) quark model couplings (dot-dashed line,
   green online) are shown as a reference.  The nucleonic coupling
   constants correspond to the DD-ME2 parametrization~\cite{ring2005};
   the hyperon--vector-meson couplings are fixed as explained in the
   text.  In panel (a) we assume $x_{\Lambda\sigma} = 0.58$, as in the
   NSC potential model, and a range $0.26\le x_{\Sigma\sigma} \le
   0.66$ which generates the shaded area; in (b) we assume
   $x_{\Sigma\sigma} = 0.448$, as in the NSC potential model, and a
   range $0.26\le x_{\Lambda\sigma} \le 0.66$. The cases $
   x_{\Sigma\sigma} = 0.46$ (left panel) and $ x_{\Lambda\sigma} =
   0.59$ (right panel), shown by dash-double-dotted (black online)
   lines, fit the depth of the potentials of the $\Sigma^-$ and
   $\Lambda$ hyperons in nuclear matter at saturation.  }
 \label{fig:hyperons_T_0}
\end{figure}

The left panel of Fig.~\ref{fig:hyperons_T_0} shows the dependence of
the EoS of hypernuclear matter on variations of the coupling constant
in the range $0.26\le x_{\Sigma\sigma}\le 0.66 $ at
fixed $x_{\Lambda\sigma}=0.58$ (a value that
corresponds to the NSC model), where 
$x_{H\sigma} = g_{H\sigma}/g_{N\sigma}$ is the ratio 
of the hyperon-$\sigma$ and the nucleon-$\sigma$ coupling constants. 
The resulting EoS covers the shaded
area, which is bound by the two EoSs corresponding to the limiting
values of the parameter $g_{\Sigma\sigma}$; in addition we show
the special case where the $g_{\Sigma\sigma}$ parameter is
fitted to reproduce the empirical value of the $\Sigma$-hyperon
potential in nuclear matter at saturation $U_\Sigma = 30$ MeV
according to 
\be \label{eq:potential}
U_\Sigma =
\frac{g_{\Sigma\sigma}}{g_{N\sigma}}
\langle\sigma\rangle+\frac{g_{\Sigma\omega}}{g_{N\omega}}\langle\omega\rangle,
\ee 
where $\langle\dots \rangle$ refers to the mean-field value of the
field. The right panel of Fig.~\ref{fig:hyperons_T_0} shows the
dependence of the EoS of hypernuclear matter on variations of the
$g_{\Lambda\sigma}$ coupling constant in the range $0.52\le
x_{\Lambda\sigma}\le 0.66 $ at fixed
$x_{\Sigma\sigma}=0.448$ (as implied by the NSC
model). The set of resulting EoS covers the shaded area; we also show
explicitly the limiting cases as well as the special case, where the
$g_{\Lambda\sigma}$ coupling is fitted to reproduce the
$\Lambda$-hyperon potential in nuclear matter at saturation $U_\Lambda
= -28$~MeV according to an equation analogous to Eq.~(\ref{eq:potential}).
As expected, the reduction of the hyperon couplings stiffens the EoS.
The stiffest hypernuclear EoS in this set corresponds to the
values of couplings $x_{\Sigma\sigma}=0.448$ and
$x_{\Lambda\sigma}=0.52$.

\begin{figure}[tbh]
 \centering
\vspace{0.5cm}
 \includegraphics[width=8.5cm]{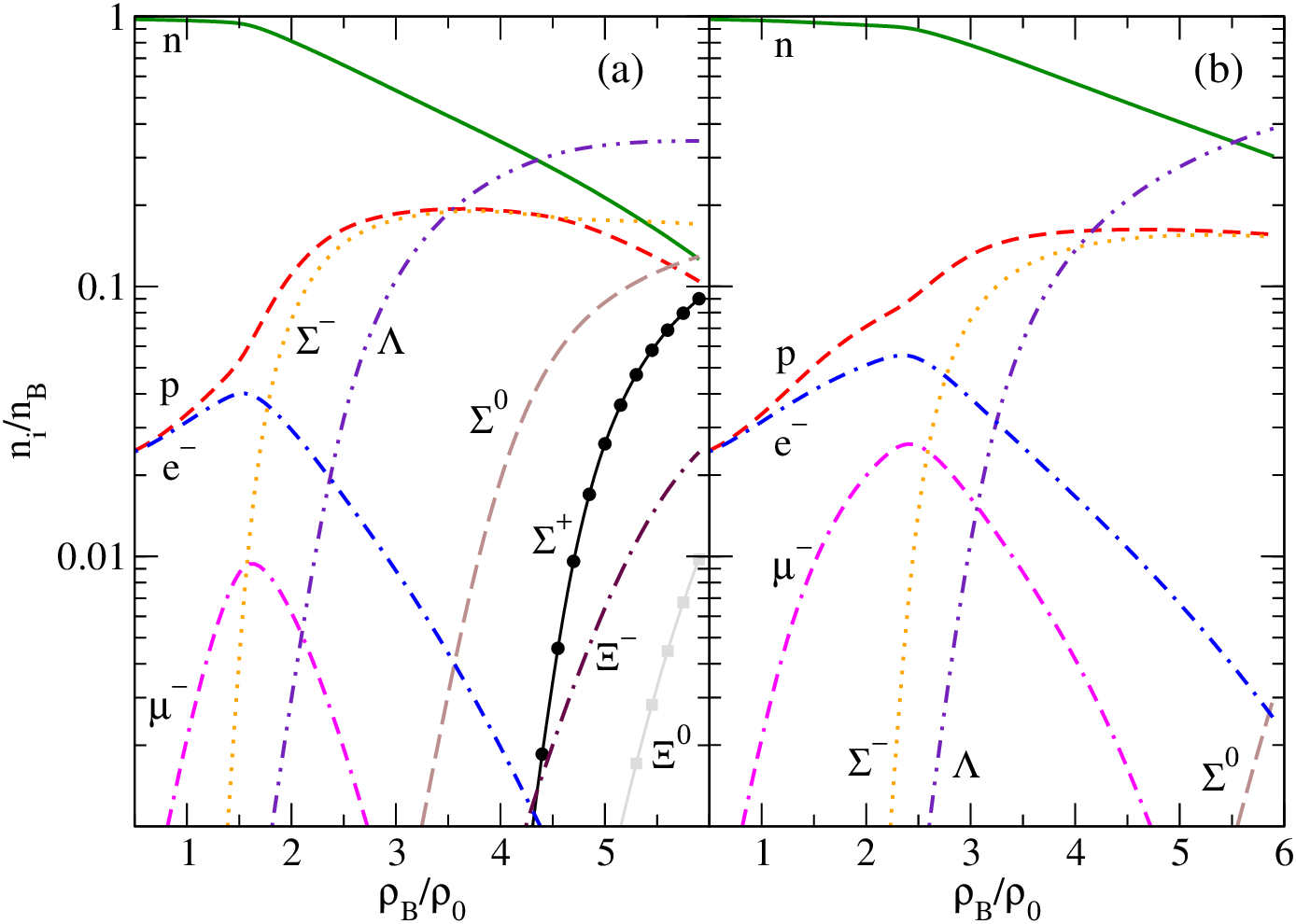}
 \caption{ (Color online)
Particle fractions in hypernuclear matter at $T=0 $: (a)
 hyperon--scalar meson couplings are fixed as in the SU(6)
   symmetric quark model,  and (b) a stiff hypernuclear EoS from
   our parameter space with $x_{\Sigma\sigma} = 0.448$ and
   $x_{\Lambda\sigma} = 0.52$.}
 \label{fig:fractions_T_0}
\end{figure}

Figure~\ref{fig:fractions_T_0} shows the particle fractions of fermions,
defined as the ratio of particle number densities to the baryon number
density for the limiting cases of the SU(6) quark model couplings and
the stiffest hypernuclear EoS. There are a number of features that are
common to both EoSs. The appearance of $\Sigma^{-}$ hyperons, which
compensate for the positive change of the nuclear matter and excess 
negative isospin of the neutrons, is favored at low density. They
contribute to the suppression and eventual extinction of electron and
muon populations at large densities, i.e., the matter becomes
deleptonized. Once the population of $\Sigma^{-}$ hyperons reaches
that of the protons and charge neutrality is established, its further
increase is not favored. Instead, an excess of neutral $\Lambda$
hyperons builds up and dominates the hyperon population at
asymptotically large densities. 
In the case of softer EoS (left panel
of Fig.~\ref{fig:fractions_T_0}) the deleptonization and onset of
hyperons like $\Sigma^{0,+}$ and cascades $\Xi^{0,-}$ occurs at
densities lower than in the case of the stiff EoS (right panel of
Fig.~\ref{fig:fractions_T_0}). This is the consequence of the fact
that the large hyperon-meson couplings favor the formation of hyperons,
which in turn soften the EoS. Thus the SU(6) quark
model, with its large meson-hyperon couplings favors the onset of
hyperons at low densities, which produces a soft EoS. Conversely, 
the stiffest EoS, which corresponds to small values of the
hyperon-meson couplings, disfavors the onset of hyperons.

\subsection{Finite temperatures}

Compact stars become transparent to neutrinos after about a minute of
formation in a supernova explosion. Prior to the neutrino-transparency
era the star is hot, with temperatures on the order of several tens of MeV,
and neutrinos are in equilibrium with matter. The neutrino thermal
distribution is characterized by the neutrino chemical potential
$\mu_{\nu}$ and neutrino fraction $Y_{\nu} = n_{\nu}/n_B$, i.e., the
ratio of the neutrino number density to the baryon number density.
The neutrino fraction is commonly parameterized by specifying the 
lepton fraction for each flavor $Y_{L} = Y_{l} + Y_{\nu}$, where $Y_l$
is the fraction of the charged leptons of a given flavor. The relevant
to proto-neutron stars range of lepton fraction is $0.1 \le Y_{L} \le 0.4$.
\begin{figure}[tbh]
\centering
\vspace{0.5cm}
\includegraphics[width=8.5cm]{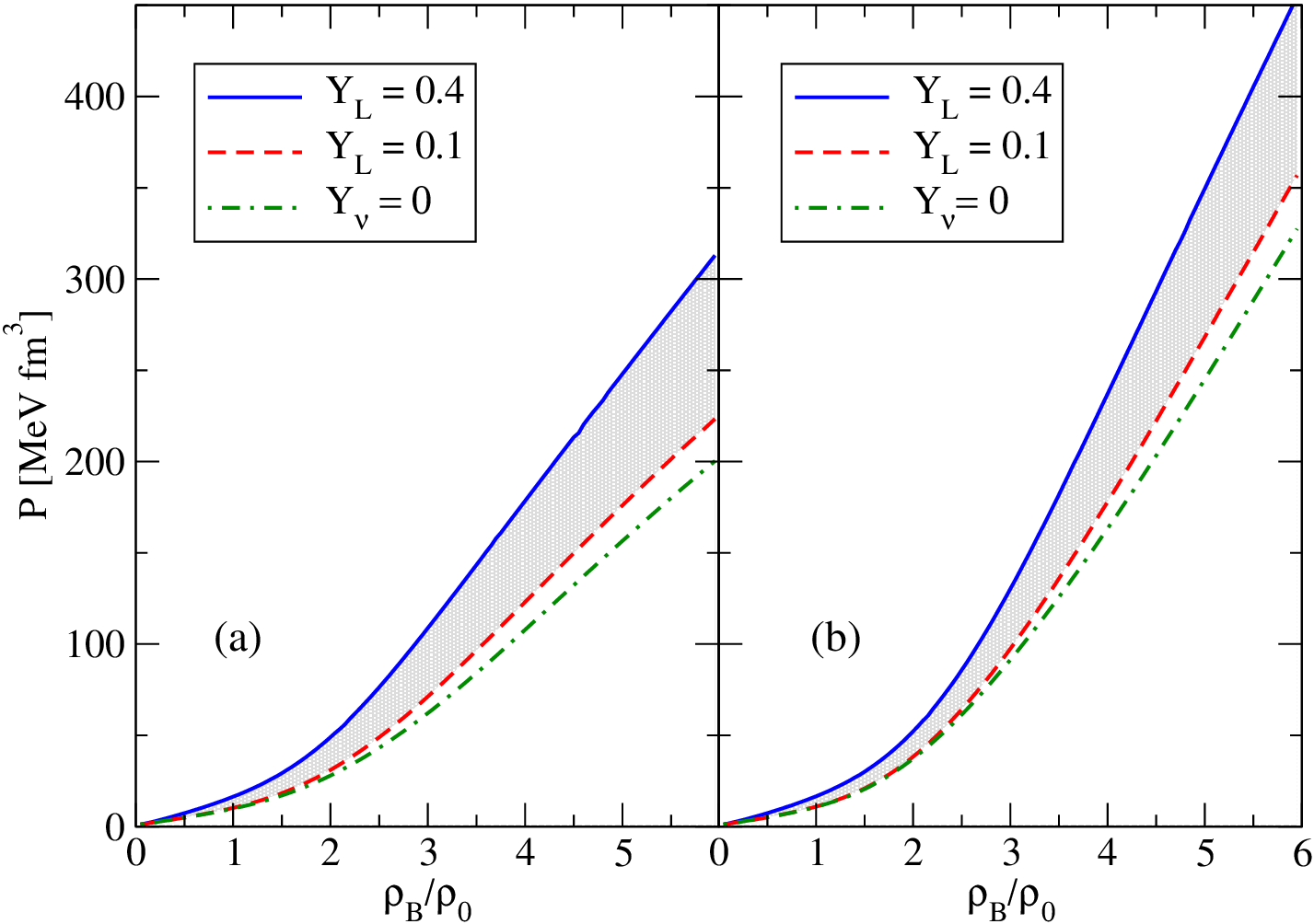}
\caption{(Color online)
Equation of state in the DD-ME2 parametrization
 \cite{ring2005} at finite temperature, $T=50$ MeV.  (a) The
 hyperon--scalar-meson coupling constants are fixed by the quark
 model. In this figure the dependence on the presence of trapped
 neutrinos is shown. The dot-dashed line (green online) corresponds
 to the case without neutrinos. The presence of
 neutrinos make the EoS stiffer.  The dashed region represents the
 variation of the EoS with the lepton fraction. The dashed line (red
 online) corresponds to a lepton fraction $Y_L = 0.1$ and the
 full line (blue online) corresponds to $Y_L = 0.4$;  (b) same as
 left, in which the coupling constants are the one of the stiffest
 case considered, corresponding to $g_{\Sigma\sigma} =
 0.448g_{N\sigma}$, fixed by the NSC potential and
 $g_{\Lambda\sigma} = 0.52g_{N\sigma}$.}
\label{fig:hyperon_T_50}
\end{figure}
Figure \ref{fig:hyperon_T_50} shows the EoS of hot hypernuclear matter
in the cases when there is a thermal population of neutrinos with $0.1
\le Y_{L} \le 0.4$ (shaded area) and in the absence of neutrinos,
i.e., $Y_{\nu} = 0$ and $Y_l$ determined from $\beta$-equilibrium with
$\mu_{\nu,l} = 0$, as in cold hypernuclear matter.  The influence of
including the neutrino population was studied for two EoS: a soft EoS
with parameters according to the SU(6) quark model and a stiff EoS with
$x_{\Sigma\sigma} = 0.448$ and $x_{\Lambda\sigma} = 0.52$ (see
Fig.~\ref{fig:hyperons_T_0} for the $T=0$ counterparts).  It can be
seen that the EoS is stiff when neutrinos are present ($Y_L ̸= 0$) and
as the fraction of neutrinos, i.e., $Y_L$, increases the EoS becomes
stiffer.  The stiffening of the EoS can be attributed to the fact that
the thermal population of neutrinos adds its contribution to the
pressure of matter. However, neutrinos stiffen the EoS also indirectly
by changing the composition of matter.
\begin{figure}[tbh]
\centering
\vspace{0.5cm}
\includegraphics[width=8.525cm]{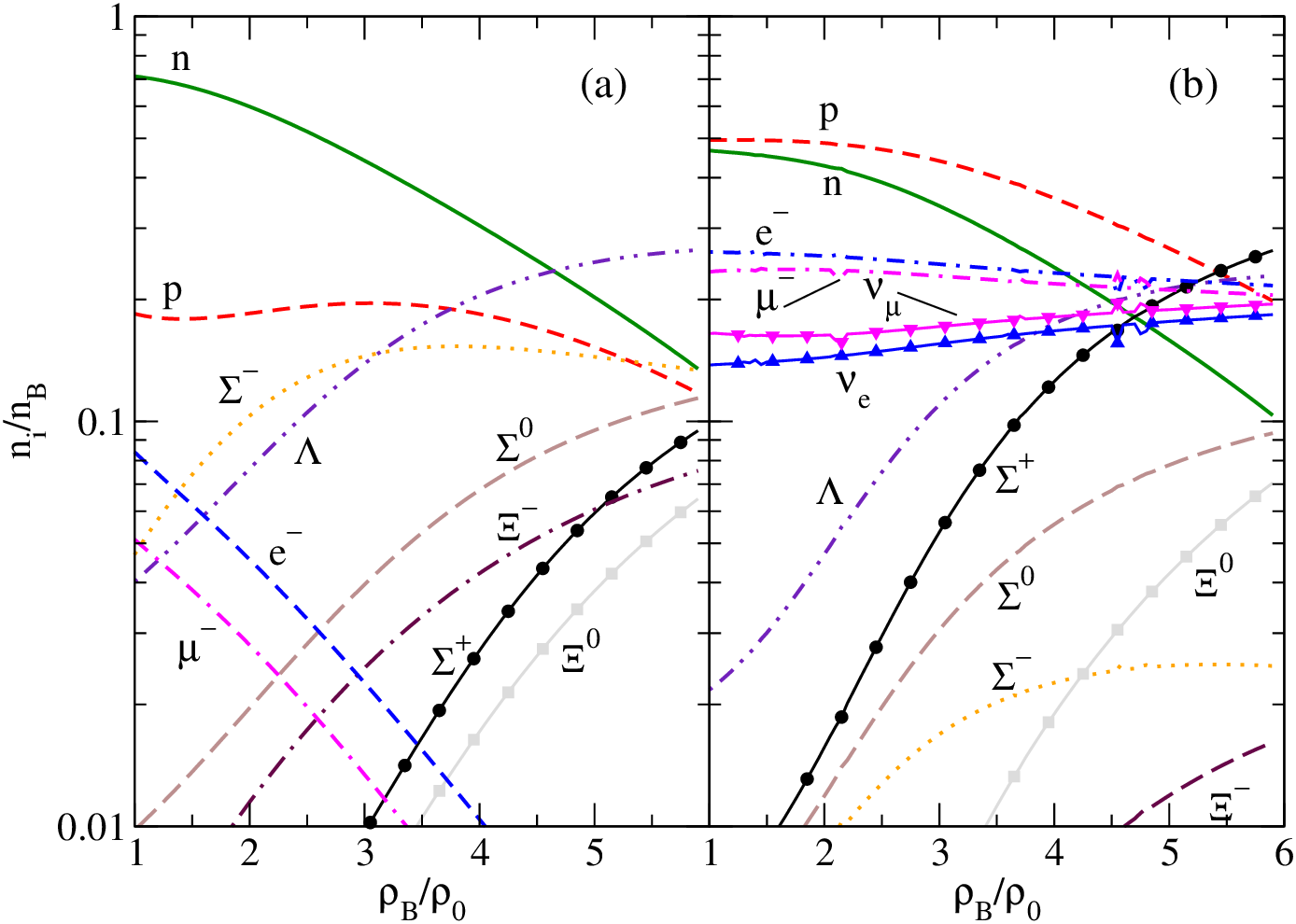}
\caption{(Color online)
  Particle fractions in hypernuclear matter at $T = 50$ MeV for soft
  hypernuclear EoS based with couplings fixed by SU(6) symmetric quark
  model. In  panel (a) $Y_\nu = 0$, and in panel (b)  $Y_L = 0.4$ 
 }
\label{fig:fractions_T_50_1}
\end{figure}
As seen in Figs.~\ref{fig:fractions_T_50_1} and
\ref{fig:fractions_T_50_2} the presence of neutrinos has a dramatic
effect on charge leptons: The deleptonization effect observed in
neutrino-less matter is reversed when neutrinos are present, which has
the consequence that the $\Sigma^{-}$ hyperons are not favored in
neutrino-rich matter. In fact the sequence of appearance of charged
$\Sigma$ hyperons is reversed (cf. the left and right panels of
Fig.~\ref{fig:fractions_T_50_1}). Since neutrinos suppress the
hyperon fractions (predominantly $\Lambda$ hyperon), at fixed baryon
density, the EoS is closer to the nucleonic one, which as we have seen,
is always stiffer than the hypernuclear EoS.  By comparing
Figs.~\ref{fig:fractions_T_50_1} and \ref{fig:fractions_T_50_2} one
can access how various choices of the hyperon--scalar-meson couplings
influences the composition of matter. The general trend, observed at
$T=0$ (Fig.~\ref{fig:fractions_T_0}), persists at non-zero
temperatures, i.e., the thresholds for the onset of the various
hyperons shift to higher densities for smaller values of the
couplings of hyperons to the scalar mesons. The models with high
values of these couplings are softer and more hyperon-rich than the ones
with smaller values of these couplings. 
\begin{figure}[b]
\centering
\vspace{0.5cm}
\includegraphics[width=8.525cm]{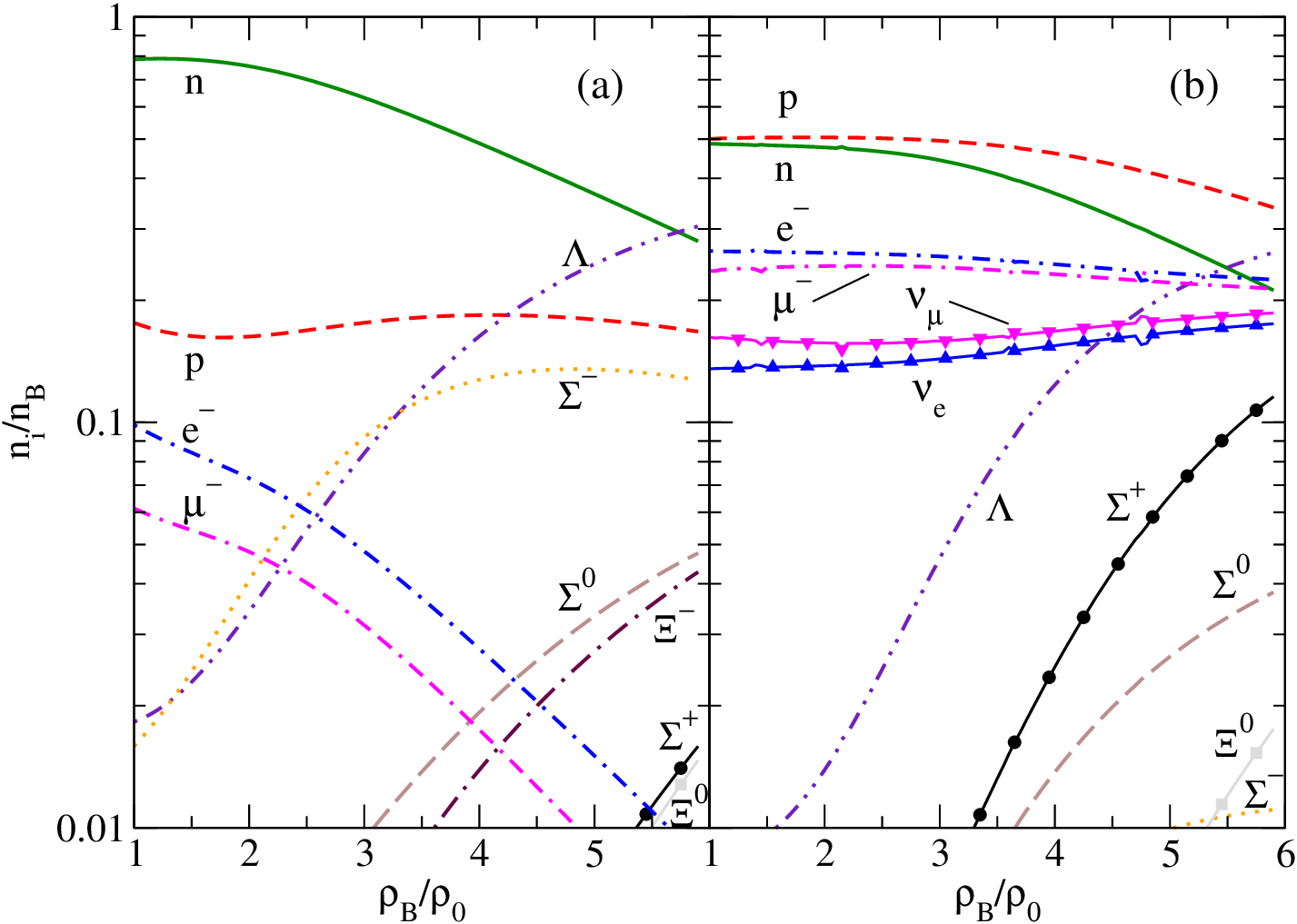}
\caption{ (Color online)
Same as in Fig.~\ref{fig:fractions_T_50_1}, 
but for a  stiff hypernuclear EoS with 
to $x_{\Sigma\sigma} = 0.448$ and $x_{\Lambda\sigma}= 0.52$. 
}
\label{fig:fractions_T_50_2}
\end{figure}

\subsection{Polytropic fits to the EoS}

Numerical relativity and astrophysics problems frequently require a
simple parametrization of the EoS matter at zero temperature. The
polytropic form of the EoS has been one of the common choices and more
recently {\it piecewise polytropic} EoS were used to construct
accurate representations of the microscopic EoS
\cite{2009PhRvD..79l4032R}.  We use below the 
representation
\be\label{eq:poly} 
P = \sum_{i=1}^4
K_i\rho^{\Gamma_i}\theta(\rho-a_i\rho_0)\theta(b_i\rho_0-\rho), 
\ee
where $\Gamma_i$ is the polytropic exponent, $K_i$ is a dimensionful
constant, and $\rho_0$ is the saturation density. These parameters are used
to fit the EoS with a given interval of densities, the lower and upper
boundaries specified by the parameters $a_i$ and $b_i$. The six
different EoSs that were parameterized using Eq.~(\ref{eq:poly}) include
a nuclear EoS, a hypernuclear EoS based on the SU(6) symmetric quark
model, and four additional EoSs with $x_{\Sigma\sigma}$ and
$x_{\Lambda\sigma}$ from the range discussed in the previous
sections. The details are given in  Table~\ref{tab:properties}.
\begin{table}[t]
 \centering
 \caption{Piecewise polytropic parametrization of various EoS
   according to Eq.~(\ref{eq:poly}).
 Set A corresponds to  a pure nucleonic EoS. The hyperonic EoS with
 different values $x_{\Sigma\sigma}$ and
 $x_{\Lambda\sigma}$ are arranged as follows: 
 set B, $x_{\Lambda\sigma} = 0.52$ and $x_{\Sigma\sigma} = 0.448$;
 set C, $x_{\Lambda\sigma} = 0.66$ and $x_{\Sigma\sigma} = 0.448$;
 set D, $x_{\Lambda\sigma} = 0.58$ and $x_{\Sigma\sigma} = 0.26$ and
 set E, $x_{\Lambda\sigma} = 0.58$ and $x_{\Sigma\sigma} = 0.66$;
 Set F  refers to the hyperonic EOS with $x_{YY\sigma}$ fixed to the
 values of the SU(6) symmetric quark model.
 The dimensionful constant $K_i$ is given in units of  MeV fm$^{3+3\Gamma_i}$ .}
\label{table:polytropes}
\vspace{0.1cm}
 \begin{tabular}{|l|cccc|cccc|}
\hline
        $i~$ &     $ K_i$   & $\Gamma_i$ & $a_i$ &$b_i$&     $ K_i$   & $\Gamma_i$ & $a_i$ &$b_i$\\
\hline
          &         & Set A &  &      &   & Set B &  &\\
\hline
    1 & 3.68535~~ & 2.86324 & 0.5  & 1.35 &3.56837 & 3.12001 & 0.5 & 2.5\\
    2 & 3.73399 & 3.07344 & 1.35 & 3   &7.66522 & 2.27431 & 2.5 & 3\\ 
    3 & 6.89825 & 2.54758 & 3    & 5   &12.0670 & 1.86718 & 3   & 4 \\
    4 & 13.2029 & 2.16247 & 5    & 10  &12.9132 & 1.81431 & 4   & 6  \\
\hline
          &         & Set C &  &      &   & Set D &  &\\
\hline
    1 & 3.76801 & 2.98366 & 0.5  & 2.25& 3.55834 & 3.12437 & 0.5 & 2.5\\
    2 & 9.15775 & 1.86209 & 2.25 & 3   & 9.18876 & 2.08451 & 2.5 & 3  \\
    3 & 9.64638 & 1.80967 & 3    & 5   & 12.9246 & 1.77419 & 3   & 5  \\
    4 & 7.79875 & 1.94152 & 5    & 6   & 15.7905 & 1.64781 & 5   & 6  \\
\hline
        &         & Set E   &  &      &   & Set F &  &\\
\hline                                 
    1 & 4.24406 & 2.50544 & 0.5  & 3.5      & 4.01772 & 2.62586 & 0.5  & 2   \\
    2 & 9.27949 & 1.88033 & 3.5  & 4.25     & 5.49227 & 2.13420 & 2    & 4.25\\
    3 & 18.9055 & 1.38903 & 4.25 & 5        & 10.5763 & 1.68309 & 4.25 & 5   \\
    4 & 25.6369 & 1.19845 & 5    & 6        & 24.7673 & 1.15400 & 5    & 6   \\
\hline
 \end{tabular}
 \label{tab:properties}
\end{table}
 The polytropic representation above should be supplemented by a
 suitable EoS of matter in the crust of a neutron star below the
 density $\rho_B/\rho_0\le 0.5$. This should be matched to given EoSs
 which are fitted only up to the crustal base, assumed to be at
 $0.5\rho_0$. To optimize the fit, the density segments were chosen to 
optimally reflect the changes in the EoS; for example one knot, in the
case of hypernuclear EoS, is chosen to be the density of the onset of
hyperons. The change of the polytropic exponent with density is
consistent with the expectation that as matter becomes more
relativistic, the polytropic exponent decreases.  

\subsection{Stellar structure}
\begin{figure}[t]
 \centering
\vspace{0.5cm}
 \includegraphics[width=8.5cm]{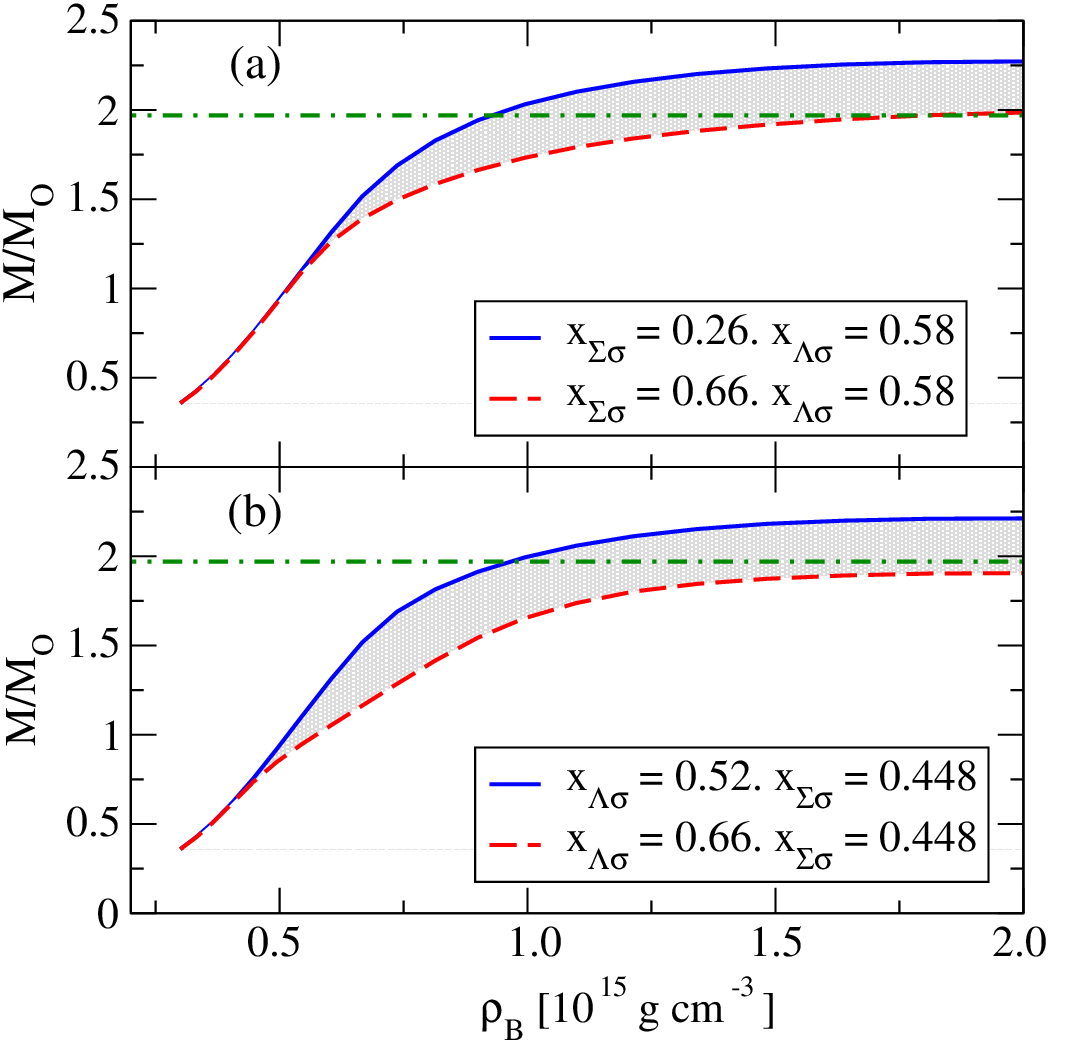}
 \caption{(Color online)
Dependence of the gravitational mass of compact hypernuclear
   stars on central density at zero temperature. The solid (blue
   online) and dashed (red online) show the limiting cases of
   parameter space as indicated in the panels (a) and (b). The dash-dotted (green online)
   line shows the observational lower limit on the maximum mass $1.97 M_{\odot}$.
} 
 \label{fig:mass_vs_rho}
\end{figure}
\begin{figure}[t]
 \centering
\vspace{0.5cm}
 \includegraphics[width=8.5cm]{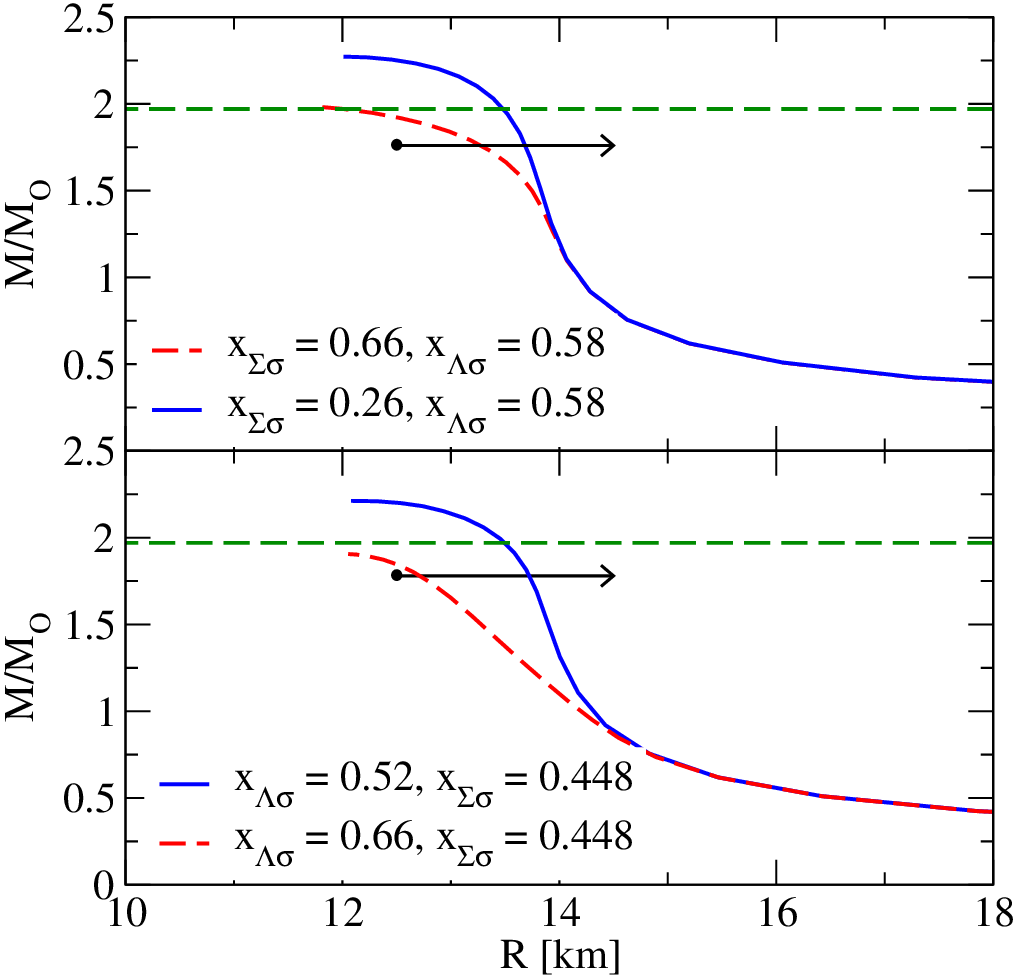}
 \caption{(Color online)
The mass--radius relations for compact hypernuclear stars at zero
temperature.  The labeling and parameter space is as in
Fig.~\ref{fig:mass_vs_rho}. The arrow shows the mass-radius constraint
of Ref.~\cite{2013ApJ...762...96B} at 2$\sigma$  level, 
which is $M = 1.76M_{\odot}$ and $R\ge 12.5$ km.
}
 \label{fig:mass_vs_radius}
\end{figure}
The spherically symmetric solutions of Einstein's equations for
self-gravitating fluids are given by the well-known
Tolman-Oppenheimer-Volkoff equations, which can be integrated for any
given EoS.  It is convenient to parametrize the equilibrium sequences
of non-rotating configurations at zero temperature by their central
density. A universal feature of these solutions is the existence of a
maximum mass for any EoS; i.e., as the central density
is increased a sequence reaches the configuration with the maximum
mass and the stars with larger central densities are unstable towards
gravitational collapse. A condition of stability for a sequence of
configurations is $dM/d\rho_c \ge 0$; i.e., the mass should be an
increasing function of the central density.  Alternatively, the
stability analysis of the lowest-order harmonics of pulsation modes
(e.g., the fundamental radial pulsations) allows access to the
stability of a configuration, for these are damped for stable stars
and increase exponentially for unstable stars.

The range of considered hypernuclear EoS translates into the band of
the stable configurations shown in Fig.~\ref{fig:mass_vs_rho}.  The
hypernuclear configurations branch off from the purely nuclear
configurations once the central density of a configuration reaches the
threshold for appearance of hyperons in matter.  The gravitational
mass of hypernuclear stars increases with the density, indicating a
stable branch of these objects, and reaches the maximum mass $\le
2.25M_{\odot}$ for densities of order $7\rho_0$. Most of the sequences
generated by the parameter space of the couplings considered is
compatible with the observational bound $M/M_{\odot}\ge 1.97$. There
is also room left for larger mass stars to allow for statistical
distribution of the masses of neutron stars beyond this limit. Note
that $1.4M_{\odot}$ canonical mass stars would be pure nucleonic for
the softer subclass of EoS considered, whereas they would contain
hypernuclear matter for the harder subclass; however  all stars
with $M> 1.5M_{\odot}$ contain hypernuclear matter.

In Fig.~\ref{fig:mass_vs_radius} we show the mass-radius relationship
for the hypernuclear sequences. The configurations with masses close
to the maximum mass $M\simeq 2M_{\odot}$ have radii of $R\simeq
12$~km, whereas the canonical mass stars have radii of order
14~km. The figure also shows the bound, which predicts that PSR
J0437-4715, being a $M = 1.76M_{\odot}$ neutron star, has a radius $R
> 12.5$~km at the $2\sigma$ level. Clearly, the hypernuclear stars are
consistent with this observation.  Finally we note that it is not
excluded that the low-mass neutron stars with $M\sim 1.2M_{\odot}$ may
already contain hypernuclear matter.

\section{Conclusions}
\label{sec:6}

Despite decades of theoretical research on hypernuclear matter, the
appearance of hyperons in compact stars remains an open issue. While
the recent astrophysical measurements exclude a significant fraction
of soft EoSs, the hyperonization of dense nuclear matter remains a
serious possibility. Our present study confirms this within a
relativistic density functional approach to nuclear matter, where we
investigated the impact of variation of the hyperon--scalar-meson
couplings on the EoS of hypernuclear matter. The range of found EoS is
sufficiently stiff to produce heavy compact stars ($M\le 2.25
M_{\odot}$). The radii of our sequences are located in the range of
$12\le R\le 14$ km.  Piecewise polytropic fits for six representative
EoSs are provided, which span the complete range of EoS from our
parameter study.

The parameter space of couplings of hyperons to scalar mesons was
explored, holding density-dependent nucleonic couplings fixed to their
values suggested by the DD-ME2 parametrization of the nuclear density
functional~\cite{ring2005}. To allow for hyperonization in massive
stars a requirement is to have small ratios of the
hypernuclear-to-nuclear couplings; in particular, hyperons need to be
coupled to scalar mesons weaker than predicted by the SU(6) quark
model.

By extending our studies to nonzero temperature and including thermal
ensemble of neutrinos (present in a compact star during the first
minute after birth) we confirm that the neutrinos stiffen the
high-density EoS and impact the charge lepton content of hypernuclear
matter. Instead of deleptonization with increasing density, seen in
neutrino-less matter, the abundances of charged leptons remain
constant, which has the consequence that the thresholds for appearance
of charged $\Sigma$'s is reversed.

The hyperonic abundances found in our EoS are broadly consistent with
the predictions of other models, both early and recent, as these are
determined by their mass spectrum and conditions of charge neutrality
and $\beta$-equilibrium. By varying the scalar--meson couplings, we
showed that the stiffer the EoS the larger are the densities at which
any type of hyperon appears. For example, if for
soft EoS a substantial amount of cascades can be built up, only trace
amounts are found in stiff EoS. Furthermore, at non-zero temperatures
the thresholds for the onset of hyperons are located at densities
lower than those at zero temperature. It is evident that during the
early cooling stage of a neutron star, i.e., the period where the
temperature drops from tens of MeV to a MeV and matter becomes
transparent to neutrinos, a rearrangement of particle content of matter
must take place.

\section*{Acknowledgments}

We gratefully acknowledge the support by the HGS- HIRe graduate
program at Frankfurt University (G.\ C.) and a collaborative
research grant of the Volks\-wagen Foundation (A.\ S.). We thank B. Friman,
D.\ H.\ Rischke, P.\ Ring, and D.\ Vretrenar for dis\-cussions and
R.\ G.\ E.\ Timmermans and S. Typel for useful communications.

\end{document}